\input{aipcheck}

\documentclass[final] {aipproc}

\layoutstyle{6x9}

\newcommand{\td}{T_{90}}
\newcommand{\tem}{{\cal T}_{50}}
\newcommand{\flux}{F_{\rm 1s}}
\newcommand{\fl}{\cal F}
\newcommand{\Epk}{E_{\rm pk}}
\newcommand{\Epkin}{E_{\rm pk;intrinsic}}
\newcommand{\REpk}{{\cal R}E_{\rm pk}}
\newcommand{\SF}{{\cal S}_{\cal F}}
\newcommand{\lag}{\tau_{\rm lag}}

\begin{document}

\title{Factor analysis of the spectral and time behavior of long GRBs}


\author{Zsolt Bagoly}{ address={Dept. of Physics of Complex Systems, E\" otv\"
os University, H-1117 Budapest, P\'azm\'any P. s. 1/A, Hungary}} 
\author{Luis Borgonovo}{ address={ Stockholm Observatory, AlbaNova, SE-106 91
Stockholm, Sweden}}
\author{Istv\'an Horv\'ath}{ address={ Dept. of Physics, Bolyai Military
University, H-1581 Budapest, POB 15, Hungary}}
\author{Attila M\'esz\'aros}{ address={ Astronomical Institute of the Charles
University, V Hole\v{s}ovi\v{c}k\'ach 2, CZ-180 00 Prague 8, Czech Republic}}
\author{Lajos G. Bal\'azs}{ address={ Konkoly Observatory, H-1525 Budapest, POB 67, Hungary}}

\begin{abstract}
A sample of 197 long BATSE GRBs is studied statistically.
In the sample 11 variables, describing for any burst the time behavior of the
spectra and other quantities, are collected.  
The application of the factor analysis on this sample shows that five factors
describe the sample satisfactorily. Both the pseudo-redshifts coming from the
variability and the Amati-relation  in its original form are disfavored.
\end{abstract}

\maketitle


\section{Introduction}

Factor Analysis (FA) and the Principal Component Analysis (PCA) are powerful
statistical methods in the data analysis.  
\cite{bag98} showed that
the 9 variables ($T_{50}$,  $T_{90}$, $P_{64}, P_{256}$, $P_{1024}$, ${\cal
F}_1, {\cal F}_2, {\cal F}_3$ and  ${\cal F}_4$) of the BATSE GRBs 
can be satisfactorily represented by 3 hidden statistical
variables. 
\cite{borbjor06} studied the statistical properties of 197 long BATSE GRBs,
using 10 statistical variables describing the temporal and spectral properties
of GRBs.  Performing a PCA they concluded that about 70~\% of the total
variance of the parameters were explained by the first 3 Principal Components
(PCs). 

FA assumes that the observed variables can be explained as a linear combination of
hidden variables as given by:
\begin{equation} \label{eq1} {x = \Lambda f + \varepsilon}\,, \end{equation}
\noindent where ${x}$ marks an observed variable of $p$ dimension, ${\Lambda}$
is a matrix of $p \times m$ dimension ($m<p$) and $f$ means a hidden variable
of $m$ dimension. The components of ${\Lambda}$ are called loadings and those
of $f$ factor scores; $\varepsilon$ is a noise term. Observation yields ${x}$
while the quantities on right-hand-side of Eq.\ref{eq1} have to be computed by
a suitable algorithm: here we use the Maximum Likelihood (ML) method.  An
interesting property of the ${\Lambda}$ matrix of factor loadings is that,
after undertaking  an orthogonal transformation (rotation) on it, one gets an
other possible factor solution. Rotation is often useful to get a solution,
which is much easier to interpret: we use the varimax rotation in our
calculations.

\section{The sample}

Here we apply the FA on the same sample of 197 long GRBs investigated by \cite{borbjor06}: for each burst we use
the following 11 variables: {
 duration time $\td$,
 emission time $\tem$,
 autocorrelation function (ACF) half-width $\tau$,
 variability $V$,
 emission symmetry $\SF$,
 cross-correlation function time lag $\lag$,
 the ratio of peak energies $\REpk$,
 peak flux on $1024$ ms scale $F_{\rm 1s}$,
 fluence $\fl$,
 peak energy $\Epk$,
 and low frequency spectral index $\alpha$.
}

It is worth mentioning here that, similarly to \cite{borbjor06}, we do not
consider the fluence on the highest channel ($> 300$ keV) separately,
although in \cite{bag98} this variable alone defined a PC (factor).
This choice is motivated by two arguments: first, because usually the fluences
on the fourth channel are often vanishing or have great errors ("the values are
noisy"); second, as it is noted by \cite{borbjor06}, in a sample given by
long-soft GRBs {\it only}, this quantity is less important.

\section{Results and discussion}

The $m$ number of
factors, which satisfactorily reproduce the
original correlation matrix, can be constrained 
(\cite{KS73}) by the inequality of 
$ m \leq (2p+1- \sqrt{8p+1})/2 \:$, which here gives $m \leq 6.782 $. 
Since the number of factors is an integer, $m=6$ is the maximum value. 

The ML method used here gives also a probability 
of the null hypo\-thesis, i.e., 
that the correlation matrix of the observed variables and that reproduced 
by the factor solution 
are statistically identical. 
For 4, 5 and 6 factors, we get 
for the validity of the null hypo\-thesis 
the probabilities of 0.000384, 0.0872, and 0.0973, respectively.
These calculations show that 5 factors are already sufficient.
The choice of 5 factors can be supported from the cumulative variances too.

\begin{table}
\begin{tabular}{lrrrrr}
\hline
 Variable &  Fact. 1 & Fact. 2 & Fact. 3 & Fact. 4 & Fact. 5\\
\hline
$\log \td $  & {         0.52  } & {  -0.05  } & {  0.16  } & {  0.06  } & {  0.83 } \\
$\log \tem$  & {        0.84  } & { -0.04  } & {  0.01  } & {  0.34  } & {  0.37 } \\ 
$\log \tau$    & {       0.88  } & {  -0.01  } & { -0.02  } & {  0.24  } & {  0.14 } \\
$\log V $ & {  0.32  } & {   0.05  } & {  0.18  } & {   0.72  } & { -0.06 } \\ 
$\log \SF  $    & {   0.10   } & { 0.05   } & { -0.17   } & { 0.46  } & {  0.06 } \\
$\log \lag $  & {    0.24   } & {  0.02   } & { -0.49   } & { -0.28  } & {  -0.03 } \\  
$\log \REpk$   & {   -0.05   } & {  0.16  } & {  -0.11   } & { -0.49  } & {  -0.07 } \\
$\log \flux$    & {   -0.03   } & { 0.99   } & { 0.07  } & { -0.08  } & { -0.04 } \\ 
$\log \fl  $ & {    0.66   } & { 0.57   } & { 0.39  } & {  0.05  } & {  0.15 } \\
$\log \Epk $ & {     0.25   } & { 0.16   } & { 0.85  } & { -0.23  } & {  0.01 } \\ 
$\alpha     $     & {   -0.05   } & { -0.03   } & { -0.32  } & { -0.07  } & { -0.23 } \\
\hline
{\em SSload}&     2.43  & 1.37  & 1.32  & 1.30 &  0.95 \\
{\em PropVar} &  0.22  & 0.12  & 0.12  & 0.12 &  0.09 \\
{\em CVar } & 0.22  & 0.34  & 0.46  & 0.58 &  0.67 
\end{tabular}
\caption{Factors coming from FA with five factors 
{\em SSload} is the sum of the squares of the loadings; {\em PropVar} defines the proportion of 
{\em SSload} to the sum of variances of the input variables; {\em CVar} defines the sum of proportional variances.
}
\end{table}

The {\bf first factor} is defined by $\tau,\, \tem,\, \td$, and $\fl$, i.e., {the
first factor is given mainly by the temporal properties}. 
The measures $\tau$ and $\tem$ are preferred over $\td$.

The {\bf second factor} is given mainly by $F_{\rm 1s}$ and $\fl$.
Hence, {the second factor is related to the observed strength of the burst}. 
The loadings of $\lag$ and $V$ are negligible, and hence there is no direct support for the
luminosity estimators based on these two variables (\cite{rafe00,re01,nor02,hak03}).

The first two factors are in accordance with \cite{bag98}
claiming that - among fluence, peak flux and duration - two principal
components or factors exist.

{The {\bf third factor} is mainly driven by $\Epk$}. 
It is interesting that this peak energy (break-energy) in
the spectra is appearing so dominantly as a significant variable in the
third factor. 
It emphasizes that the spectrum itself is an important quantity
(a trivially expectable result), and in the spectrum  $\Epk$ itself is a
significant descriptor (this is not a triviality).
The loadings of $\log \lag$ and $\log \fl$ 
are also important in the third factor. It means 
that there is some connection of
$\lag$ with the emitted energies of GRBs, and thus the luminosity
indicator based on the spectral lag seems to be indirectly
supported  by the structure of third factor (\cite{nor02,ry05a,ry05b}).
If the Amati-relation (\cite{am02}) stands then there should be a linear
connection between $\log \Epkin$ and $\log E_{iso}$.
The Amati-relation was indeed predicted by the strong correlation between $\log \fl$ and $\log \Epk$ (\cite{Lloyd00}).
The correlation between $\log \fl$ and $\log \Epk$ does not mean that
there is a linear connection {\it only} between $\log E_{iso}$ and $\log \Epkin$. 
In fact, 
\cite{borbjor06} arrived also to the conclusion that 
\begin{equation}
\log E_{iso} = a_1 \log \Epkin  + b_1 \log \tau_{intrinsic} + c_1
\end{equation}
should hold with some suitable $a_1, b_1, c_1$ constants
($\tau_{intrinsic} =\tau/(1+z)$). 
Note that $\tem$ and $\tau$ strongly correlates with each other, i.e., in this equation
either $\tau_{intr}$ or ${\cal T}_{50;intrinsic}$  can be used.
Recently, the validity of the Amati-relation has been a matter of intensive discussion (\cite{ry05b,napi05,Fir06,but07}).
The factor loadings show that $\log \fl$ is
explained basically by the first three factors. Since Fact. 1 is mainly
given by $\log \tau$, Fact. 2 by $\log \flux$ and Fact. 3 by
$\log \Epk$, all this suggests that a relation of the form
\begin{equation}
\label{epkinmod}
\log E_{iso} = a_2 \log \Epkin + b_2 \log \tau_{intr} + c_2 \log L_{iso} +d
\end{equation}
should exist, with some suitable $a_2, b_2, c_2, d$ constants.
Note that a similar relation was proposed also by \cite{Fir06}. 
It follows from 
the first three factors that the relationship of
$\log \fl$ and $\log \Epk$ is less important than that of the variables
dominating Fact. 1 and 2, because $\log \fl$ and $\log \Epk$ together
are mainly determined by Fact. 3, and thus for their relation one cannot omit
the variables that are dominating Fact. 1 and Fact. 2, respectively. This
fact disfavors a simple linear relationship {\it only} between  $\log \Epkin$
and $\log E_{iso}$. 

{The {\bf fourth factor} is dominated by $V$, $\SF$ and  $\REpk$}. 
However, according to \cite{rafe00} and \cite{re01}, the variability should be
coupled to the luminosities of GRBs, and hence to the fluence and peak flux. No
such connection is supported by the fourth factor. Hence, some queries emerge
here for the redshift estimations derived from the variability.

{The {\bf fifth factor} is dominated by $\td$ and $\tem$}.  
This shows that $\td$ and $\tem$ are
not completely equivalent, though $\tem$ better characterizes a burst.

\section{Conclusions}

The results of the paper may be summarized as follows.

\begin{itemize}
\item No more than 5 factors should be introduced. This
essential lowering of the significant variables is the key result of
this paper.
\item The structure of factors is similar to the PCs of
\cite{borbjor06}.  The number of important quantities is more accurately
defined here.
\item The first factor is given mainly by the temporal variables, and the
quantities $\tau$ and $\tem$ are preferred.
\item The second factor is related to the strength of the burst.
\item The connection of $\Epk$ in the third factor with other
quantities, and the structure of the first three factors cast
considerable doubts about the Amati-relation in its original form.
For the luminosity indicators based on the spectral lag some support
emerges from the third factor.
\item The variability in fourth factor does not support its connection
to the intrinsic luminosities, and
the pseudo-redshift estimations based on the variability.
\item The fifth factor shows that $\td$ and $\tem$ are not completely
equivalent.
\end{itemize}

Because all these conclusions are obtained from the measured data alone, all
models of long GRBs must respect these expectations.


\begin{theacknowledgments}
Thanks are due to the valuable discussions with Claes-Ingvar Bj\"ornsson,
Stefan Larsson, Peter M\'esz\'aros, Felix Ryde and P\'eter Veres.  This study
was supported by the Hungarian OTKA grant No. T48870 and 75072, 
by a Research Program
MSM0021620860 of the Ministry of Education of Czech Republic, by a GAUK grant
No. 46307, and by a grant from the Swedish Wenner-Gren Foundations (A.M.).
\end{theacknowledgments}

\bibliographystyle{aipproc}   

\end{document}